# Surface step induced bending in two-dimensional oxide nanosheet

Jiake Wei[1, 2, 3*], Masato Yoshiya[4, 5], Feng Bin[1], Naoya Shibata[1, 5], Yuichi Ikuhara[1,2, 5*]

[1] Institute of Engineering Innovation, The University of Tokyo, Tokyo 113-8656, Japan.

[2] Center for Elements Strategy Initiative for Structural Materials, Kyoto University, Kyoto 606-8501, Japan.

[3] State Key Laboratory of Catalysis, Dalian Institute of Chemical Physics, Chinese Academy of Sciences, Dalian 116-023, China.

[4] Division of Materials and Manufacturing Science, Osaka University, Osaka 565-0871, Japan.

[5] Nanostructures Research Laboratory, Japan Fine Ceramics Center, Nagoya 456-8587, Japan.

*Correspondence to: J. W. (weijiake@dicp.ac.cn), Y. I. (ikuhara@sigma.t.u-tokyo.ac.jp)

The surface steps play dramatic roles in surface dominated processes and the related properties, in which these roles would be significantly enhanced for the low-dimensional nano or quantum materials since the ratio of surface atoms is largely increased. However, the atomic structures of surface steps in the nano/quantum systems are poorly understood because it is extremely difficult to controllably introduce these defects experimentally. Here, by utilizing the focused sub-Angstrom high energy electron beam induced atomic precise etching in a scanning transmission electron microscope, we fabricate MgO two-dimensional (100) nanosheet as thin as 2 atomic layers, and controllably introduced the surface steps in the nanosheet. It is found that the surface steps could bend the MgO nanosheet and such step-induced bending becomes more pronounced for the

**nanosheet with less layer numbers. We reveal that the bending of the nanosheets is originated from the step induced step force and the reduced bending stiffness in the nanosheet system.**

The steps are the most common defects at a solid surface[1]. Atoms at a surface step usually have smaller coordination numbers compared with those in the perfect surface and the bulk. Such coordinative unsaturation would introduce further relaxation of the atoms near the step and modify the structures and properties of the surface. These effects make the steps play dramatic roles in surface dominated processes such as heterogeneous catalysis[2,3], film growth[4-6] and dislocation nucleation[7,8]. Moreover, the roles of surface and surface steps are significantly enhanced and the defects dominated novel properties could emerge when the size of the materials is reduced to the nano or quantum scales, because the ratio of the atoms at these defects is largely increased[9]. Although the atomic structures of the steps at a solid surfaces have been widely studied by both theoretical and experimental approaches[10], little is known about how the steps affect the surface structures in nano or quantum systems, such as in the two-dimensional nanosheets. This is because it is experimentally difficult to controllably introduce the atomic defects in the low dimensional structures.

Recently, electron beam irradiation in a scanning/transmission electron microscope (S/TEM) has been successfully applied to tailor materials and introduce atomic defects[11-14]. The energetic electron beam could kick out atoms in materials by the elastic and/or inelastic beam-sample scattering. By combining with the focused electron beam with sub-angstrom diameter in an aberration-corrected STEM, the atomic precision fabrication could even be achieved[14]. It has been demonstrated that the electron beam irradiation could top-down etch two-dimensional materials, like graphene and monolayer transition-metal dichalcogenide, to one-dimensional nanoribbons or nanowires[12,13]. However, for more general three-dimensional materials, atomic-scale fabrication by the electron beam is rarely reported.

Here, we demonstrated an atomic precision etching of MgO (with a thickness of ~15 nm) by the electron beam irradiation in an aberration corrected STEM. By this method, the MgO (100) nanosheets, as low as 2 atomic layers with and without surface steps, were controllably fabricated. The atomic structures of these nanosheets were investigated by atomic resolution STEM imaging. It is found that the existence of surfaces steps could bend the MgO nanosheets. Such bending is pronounced when the layer number is reduced because the bending stiffness is decreased accordingly.

MgO is a typical metal oxide with a rock-salt structure, which are widely used as the substrate for thin film growth[15] and catalyst[16-20]. The steps at the MgO surfaces are reported to be the catalytic active sites for the many chemical reactions[21]. Meanwhile, the MgO nanosheet or film was reported to be a better catalyst than the bulk counterpart[21-23], although it is still experimentally difficult to introduce such two-dimensional structures in oxide systems. In this study, we try to fabricate the MgO nanosheet by using the high energy electron beam induced etching in the STEM. To effectively etch MgO, we make use of the 80 keV focused incident electron beam in the STEM. For insulating oxide like MgO, the 80 keV electron beam could introduce significant damages by the inelastic beam-sample interaction such as the beam induced electric field effect[24]. The current density of the beam is estimated to be ~$1.4\times10^8$ Am$^{-2}$, under which the MgO was considered to lose mass stoichiometric without formation of a secondary phase like the metallic Mg[25]. Meanwhile, to reduce the undesired beam damage, the current density was decreased to ~$3.5\times10^4$ Am$^{-2}$ during the imaging the MgO nanostructures.

Figs. 1a-l and Movie S1 show the experimental demonstration of etching MgO with atomic precision by the 80 keV electron probe. Fig. 1a shows the high angle annular dark field (HAADF)-STEM image of the pristine MgO viewing from <100> direction, in which the Mg and O ions are overlapping and indistinguishable. The sample thickness is estimated to be around 15 nm by the electron energy loss spectroscopy. After focusing the electron probe on the white spot in Fig. 1a for 10 seconds, the atoms within the dotted green box in Fig. 1a, corresponding to 3 times 3 atomic columns, were kicked out by the high energy beam shown in Fig. 1b. It is important to note that a square hole with four nonpolar (001) surfaces of MgO was created after electron beam irradiation, which may be because the (001) surface is the most stable surface in MgO. Such result agrees well with the previous reports[25]. After a hole was created in Fig. 1b, the electron probe was again focused on the white spot for 3 seconds, after which, 2 times 3 atomic columns within the green box in Fig. 1b were kicked out as shown in Fig. 1c. Meanwhile, there are three atomic columns within the dotted white box in Fig. 1c appeared on the opposite surface from the focused beam. This fact indicates that some of the atoms within the yellow box of Fig. 1b were diffused to white box of Fig. 1c, which partly healed the hole created. Such electron beam

irradiation induced etching was repeated so that the rectangle hole was controlled to grow layer-by-layer. In Fig. 1l, a hole of 4 times 28 atomic columns in size was created.

It is important to note that the spatial resolution of the etching process is supposed to depend on the probe size of the incident beam, the beam broadening effect due to the thickness of the sample, and the sample drift. In the present case, although the probe size is below 1.5 Å, the sample drift and beam broadening limited the spatial resolution of the etching and the smallest hole could be fabricated is with 2×2 atomic columns (6×6Å$^2$ in sizes). Nevertheless, atomic precision etching could be realized during widening the holes in Figs. 1g-1l by carefully locating the electron beam slightly off the etched atoms from the surface as shown as the white dots in the Fig. 1g-1k.

By utilizing the electron beam irradiation induced etching, we fabricated MgO nanosheets shown in Fig. 2 and Movie S2. By focusing the electron beam at selected regions, two rectangle holes were drilled side-by-side, and thereby the MgO nanosheet was formed in between the two holes as shown in Movie S2. By further layer-by-layer etching the nanosheet, its layer numbers can be step-by-step reduced to as low as 2 layers, corresponding to half unit cells of MgO. These nanosheets maintain the structure of the bulk MgO, which is consistent with previous theoretical calculations[26,27]. Benefit from the atomic precision fabrication, the surface steps could also be introduced on the nanosheets as shown in Figs. 3a-f. It is obviously that the existences of the surface steps further relaxed the local atomic structures of the nanosheets. The nanosheets are slightly bended around the steps, in which such bending effect is more pronounced when the layer number of the nanosheet is reduced. To better visualize the bending of the nanosheet in Fig. 3g, we calculated the angles between the neighboring atomic columns with the horizontal line of the two plus half layers nanosheet in Fig. 3f. The definition of the sign of the angles was given in the left bottom of Fig. 3g. It is clearly shown that at the left and right side of the step, the angles are mainly negative (shown as green color) and positive (red color) respectively, which suggest the nanosheet bends downward at the step.

To understand the physical origin of the surface step induced bending, we measured the atomic displacements around the steps in Fig. 4. Figs. 4a-b shows the displacements around a step in (001) surface of the bulk MgO. To obtain the displacements, we firstly located positions of the atomic column in a perfect (100) surface of MgO by two-dimensional Gaussian fitting of the intensity in the HAADF image (Fig. 4a). Then the atoms in the yellow box in Fig. 4a was etched so that a surface step was introduced. The column positions around the step were located again as shown in Fig. 4b. By comparing the column positions in Fig. 4a and Fig. 4b, the displacements of the atomic columns around the surface step were measured and shown as arrows in Fig. 4a. These arrows represent how the atoms are displaced from the positions without step to the one with step. As shown in Fig. 4a, the largest displacements occur at the edge and corner of the step, in which both of the two columns are displaced downward. These results are qualitatively consistent with the previous experimental and theoretical results[10,28]. In the same way, Figs. 4c-d give the displacements around a step of the six-layer nanosheet plus a step. Similar to the case in Fig. 4a, the atoms at step edge and corner are both displaced downward, whereas these displacements have much larger sizes than the step at the bulk surface. In addition, the step induced displacements are extended to atoms far from the step and even the atoms at the farthest layer to the step are displaced downward. Such downward displacements contribute to the bending of the nanosheet. Furthermore, when the layer number of the nanosheet is reduced to 2 layers plus the step in Fig. 4e-f, the displacements become larger and shows a more obvious bending.

In the classical elastic surface theory, a surface step could be considered as a pair of dipole point forces applied on the step edge and corner respectively, which deform the crystal and induce the displacements of atoms near the step[1,29,30]. According to the displacement field of the step on the (100) surface of MgO in Fig. 4, the point forces could be marked as the red arrows in Fig. 5a, in which the directions of the forces on the step edge and corner are along the displacements on each column. It is important to note that the point forces result a total net force perpendicular to the surface. This force is the reason for the bending of the MgO nanosheet. In addition, it has been reported that the bending stiffnesses in two-dimensional materials are closely related to their layer numbers[31]. The bending stiffnesses of few-layer graphene are decreased with the reduced layer numbers[31]. It is therefore reasonable to deduce that the layer number dependent bending of MgO nanosheet is because of the change of the bending stiffnesses.

To quantitively describe the bending of MgO nanosheets and their bending stiffnesses, Fig. 5b summarized the deflections of MgO nanosheet with different layer numbers, in which the deflection is defined as the largest vertical bending distance for the columns in the layer farthest to the step as schematically shown in Fig. 5a. It is shown the deflections increase from ~20 pm to ~110 pm when the layer numbers are decreased from 8 layer to 2 layer plus a step. Since the bending stiffnesses is proportional to the reciprocal of the deflections, Fig. 5c rearranged the measured deflections in Fig. 5b as the reciprocal of deflections for different layer numbers. It is shown that the bending stiffnesses are reduced with less layer numbers of MgO (100) nanosheet, which is consistent to the system of the few-layer graphene[31].

In summary, we developed an atomic precision etching of MgO by the focused high energy electron beam irradiation in a STEM. By utilizing such etching method, the MgO (100) nanosheets as low as 2 layers, with and without surface steps were fabricated. We demonstrated the surface step induced bending of the nanosheets, which are originated from the surface step forces and the reduced stiffnesses in the nanosheet systems. These results not only developed an atomic precision fabrication strategy by the focused electron beam irradiation in an aberration corrected STEM, but also demonstrated an atomistic insight of how the defects changing the structures of materials in the nano or quantum systems. The reduced stiffnesses of MgO nanosheet makes the nanosheets extremely flexible, so that the roles of steps might be significantly enhanced, for example when interacting with the external forces and the dislocation nucleation. The step-induced bending of the nanosheet may change the localized electronic structures, which may impact the catalytic properties of MgO. Moreover, the atomic precision fabricating method developed here could further utilized to fabricate the nano/quantum structures and explore new novel properties.

**Methods:**

The MgO TEM specimens were prepared by mechanical polishing and Ar-ion beam milling. The typical sample thickness was estimated to be 15-20 nm. The atomic structures MgO were observed using STEM (ARM200CF, JEOL Co. Ltd) operated at 80 kV, with an electron probe current of ~3 pA. The convergence semi-angle of the electron probe was 30 mrad and the scattered electrons were collected using an annular detector spanning the range 90 to 200 mrad for HAADF imaging. To obtain high signal-to-noise ration images and get rid of the sample drift, all the images here were acquired by aligning 5 rapidly recoded 512×512 pixel frames with a 5μs/pixel. The images here were acquired on the [001] zone axis of MgO. To etching the the MgO, 80 kV electron beam with a probe current of ~30 pA was controlled focused and moved on the MgO sample as shown in Fig. 1. To fabricate the MgO nanosheets, the electron beam was focused at selected regions so that two rectangle holes were drilled side-by-side, and thereby the MgO nanosheet was formed in between the two holes as shown in Movie S2.

**Acknowledgments:** We thank Dr. R. Ishikawa from the University of Tokyo for useful discussions.

**Funding:** This work was supported by Grant-in-Aid for Specially Promoted Research (Grant No. JP17H06094), Grant-in-Aid for Scientific Research on Innovative Areas (Grant No. JP19H05788) from the Japan Society for the Promotion of Science (JSPS), the Elements Strategy Initiative for Structural Materials (ESISM) (Grant No. JPMXP0112101000) of the Ministry of Education, Culture, Sports, Science and

Technology (MEXT). A part of this work was conducted at the Advanced Characterization Nanotechnology Platform of the University of Tokyo and supported by the "Nanotechnology Platform" of the Ministry of Education, Culture, Sports, Science and Technology (MEXT), Japan (Grant No. JPMXP09A20UT0146).

**Author contributions:** J. W. designed and carried out the experiment, analyzed the data and wrote the manuscript. M. Y. carried out the calculations. B. F., N. S., and Y. I. discussed the results. Y. I. supervised the work. All author commented on the manuscript.

**Competing interests:** All authors declare no competing interests.

**Figures and Figure captions**

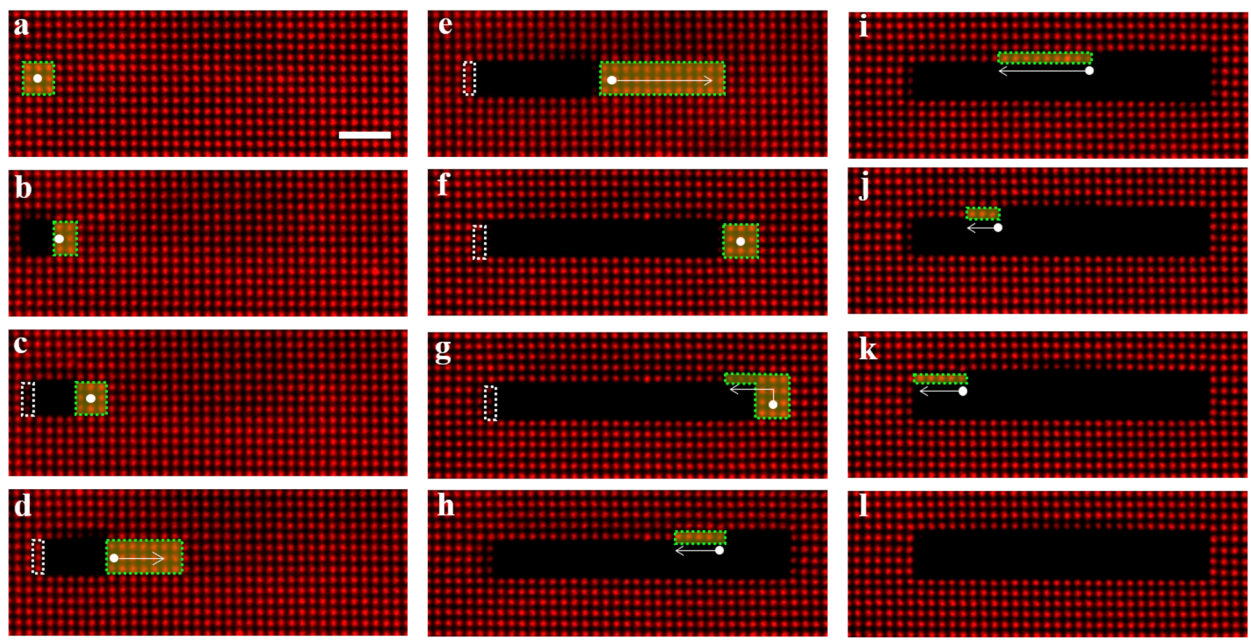

**Figure. 1 Atomic precision etching of MgO by electron beam irradiation. a-i**, Experimental images of step-by-step etching of MgO. The electron beam was focused at the white dots and was controlled to move along the white arrows. The atoms in the dotted green box were removed by the beam and the atoms within the dotted white box were healed. The scale bar is 1 nm.

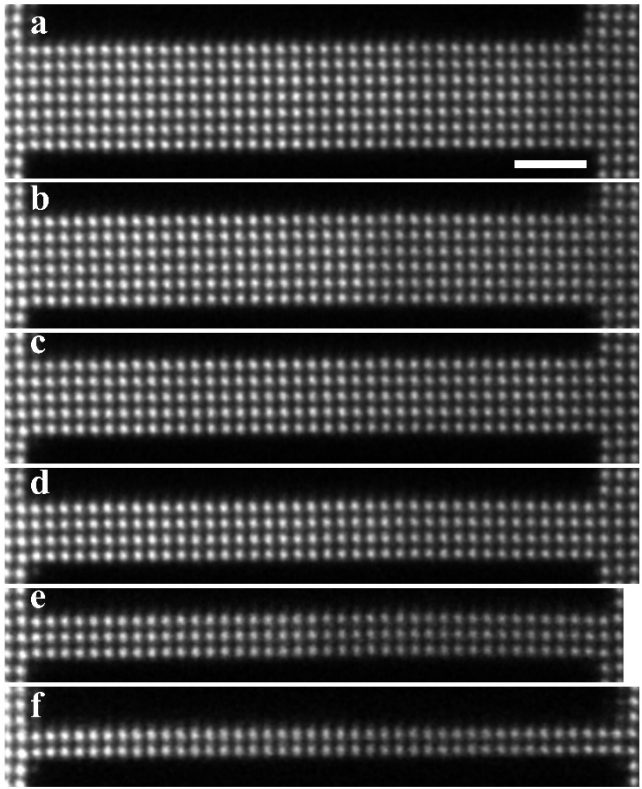

**Figure. 2 Fabricating of MgO nanosheets.** The MgO nanosheets with 7 to 2 layers were fabricated from **(a)** to **(f)**. The scale bar is 1 nm.

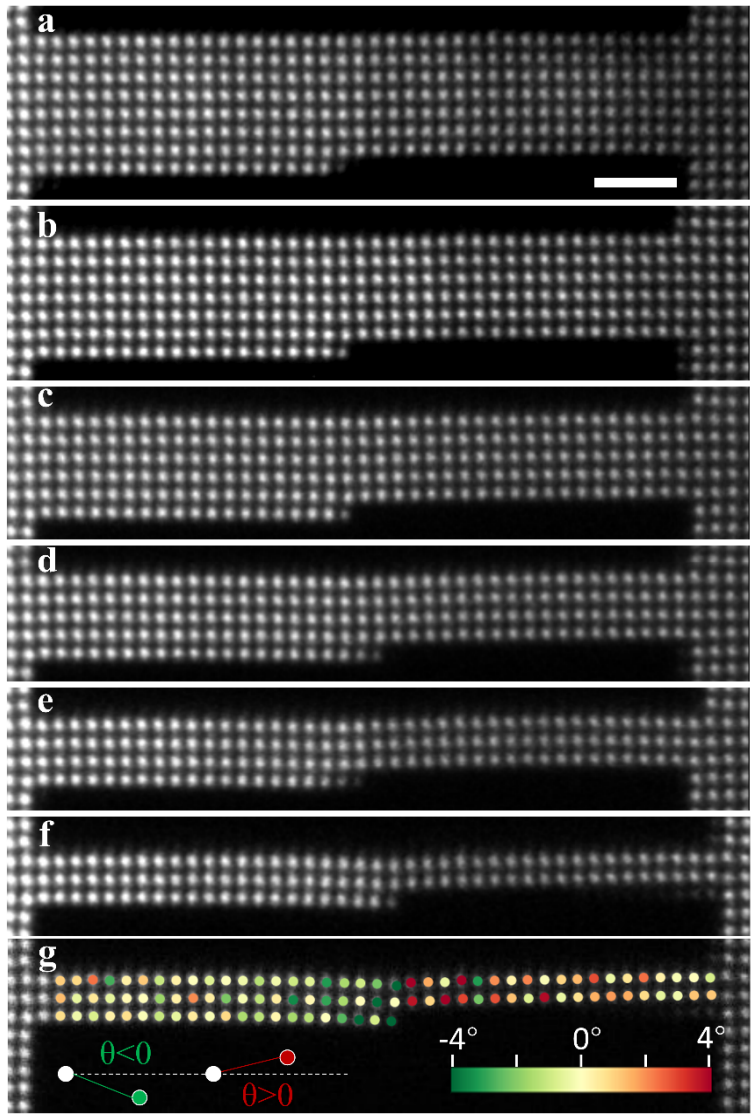

**Figure. 3 Fabricating of MgO nanosheets with surface step. a-f,** MgO nanosheets from 7 to 2 layers plus a surface step introduced. The scale bar is 1 nm. The nanosheets are bended by the surface steps. **g,** The calculated angles between neighboring atomic columns with the horizontal line, which is schematically shown left bottom of the figure. The colors of the dots on each columns indicate the angles with the color bar given on the right bottom of the figure. On the left and right side of the step, the angles are mainly negative and positive respectively, indicating the nanosheet is bending downwards by the step.

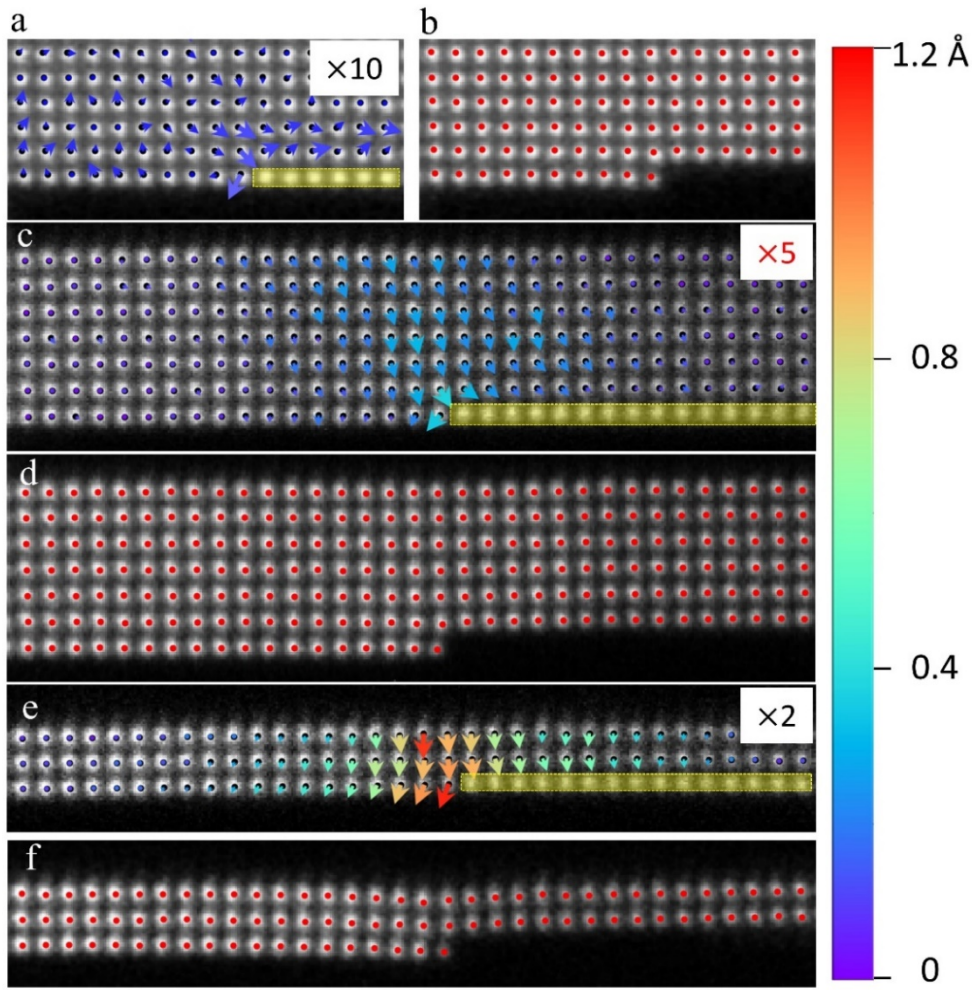

**Figure. 4 The atomic displacement around the step. a-b,** The atomic displacements of the step at bulk (100) surface. The atoms in the yellow box of (**a**) were removed by the electron beam irradiation to introduce the step in (**b**). The dots indicate the positions of atomic columns in the surface of bulk MgO with (**b**) and without (**a**) steps located by the Gaussian fitting of the intensity of each column. And the arrows in (**a**) correspond to the displacements from the atomic positions in (**a**) to the positions in (**b**). The sizes of these arrows are 10 times magnified. The color of each arrow indicates the real displacement size. **c-d,** The atomic displacements of the step in 6-layer MgO nanosheet plus a step. The sizes of the displacements (the arrows in (**c**)) are 5 times

magnified. **e-f,** The atomic displacements of the step in 2-layer MgO nanosheet plus a step. The sizes of the displacements (the arrows in (**c**)) are 2 times magnified.

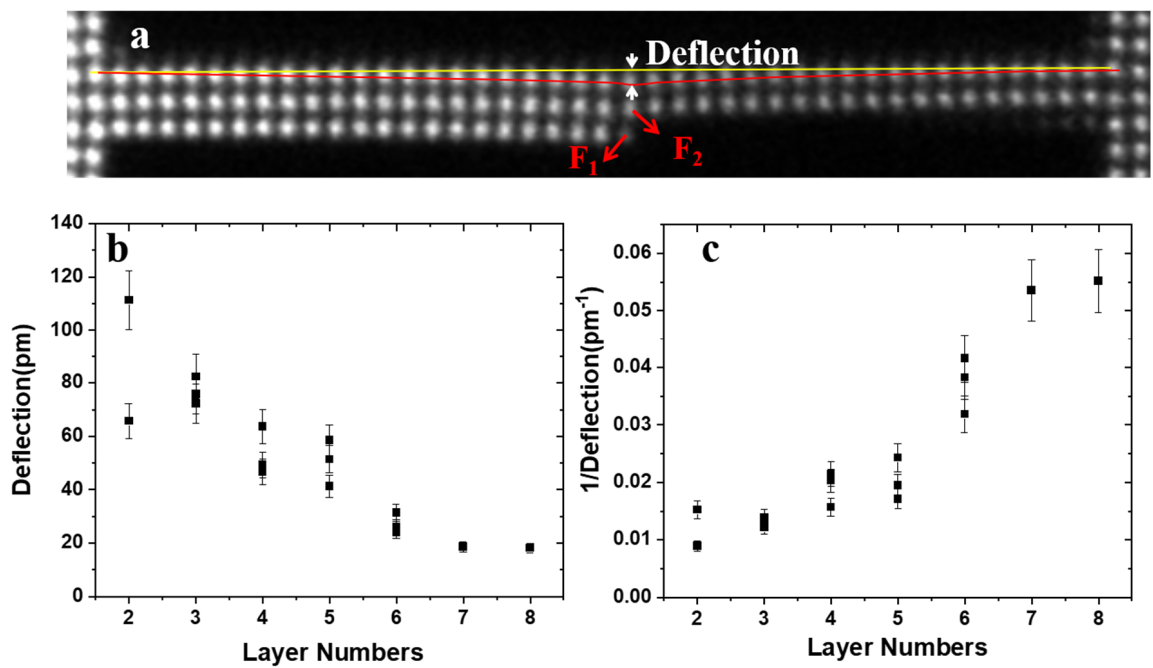

**Figure. 5 The step forces induced deflections calculated for MgO nanosheets with different layer numbers. a,** The schematically illustrate of the step forces induced deflections. The $F_1$ and $F_2$ are the point step forces applied on the corner and edge of the step, respectively. The yellow and red solid lines indicate the un-bended and bended layer of the farthest to the step, respectively. The deflection of the nanosheet is defined as the largest vertical distance of the displacements of the atoms in the layer (the distance within the two white arrows). **b,** The measured defections for different layer numbers. **c,** The reciprocal of deflections for different layer numbers. The bending stiffness is proportional to the reciprocal of deflection.

Supplementary Materials for

# Surface step induced bending in two-dimensional oxide nanosheet

Jiake Wei[1, 2, 3*], Masato Yoshiya[4, 5], Feng Bin[1], Naoya Shibata[1, 5] and Yuichi Ikuhara[1, 2, 5*]

*Correspondence to: J. W. (weijiake@dicp.ac.cn), Y. I. (ikuhara@sigma.t.u-tokyo.ac.jp)

**This file includes:**

1. Captions of Movies S1 to S2

**Other Supplementary Information for this manuscript include the following:**

Supplementary Movies S1 and S2

1. Captions of Supporting Movies

**Movie S1. Artificial stop-motion movie sequentially showing the step-by-step etching of MgO structures.** The images in Figure 1 are taken from this movie.

**Movie S2. Artificial stop-motion movie sequentially showing the fabricating of MgO nanosheet.**